\documentclass[12pt]{article}
\usepackage{epsfig}
\def\be{\begin{equation}}
\def\ee{\end{equation}}
\def\bea{\begin{eqnarray}}
\def\eea{\end{eqnarray}}
\usepackage{graphicx}

\catcode`\@=11
\def\lsim{\mathrel{\mathpalette\@versim<}}
\def\gsim{\mathrel{\mathpalette\@versim>}}
\def\@versim#1#2{\vcenter{\offinterlineskip
\ialign{$\m@th#1\hfil##\hfil$\crcr#2\crcr\sim\crcr } }}
\catcode`\@=12
\usepackage{axodraw}

\catcode`@=12
\begin{document}
\thispagestyle{empty}
\begin{flushright}
UCRHEP-T465\\
May 2009\
\end{flushright}
\vspace{0.8in}
\begin{center}
{\LARGE \bf Deciphering the Seesaw Nature of Neutrino Mass 
from Unitarity Violation\\}
\vspace{1.2in}
{\bf Ernest Ma\\}
\vspace{0.2in}
{\sl Department of Physics and Astronomy, University of California,\\
Riverside, California 92521, USA\\}
\end{center}
\vspace{1.2in}
\begin{abstract}\
If neutrino masses are obtained via the canonical seesaw mechanism, based on 
an underlying $2 \times 2$ mass matrix, unitarity violation of the neutrino 
mixing matrix is unavoidable, but its effect is extremely small.  On the other 
hand, in the $inverse$ (and $linear$) seesaw mechanisms, based on an underlying 
$3 \times 3$ mass matrix, it can be significant and possibly observable.  This 
$3 \times 3$ matrix is examined in more detail, and a {\it new} variation (the 
$lopsided$ seesaw) is proposed which has features of both mechanisms.  A 
concrete example based on $U(1)_N$ is discussed.
\end{abstract}

\newpage
\baselineskip 24pt
In the famous canonical seesaw mechanism, the standard model 
(SM) of particle interactions is implemented with a heavy singlet 
``right-handed'' neutrino $N_R$ per family, so that the otherwise massless 
left-handed neutrino $\nu_L$ gets a mass from diagonalizing the $2 \times 2$ 
mass matrix spanning 
($\bar{\nu}_L, N_R$):
\begin{equation}
{\cal M}_{\nu,N} = \pmatrix{0 & m_D \cr m_D & m_N},
\end{equation}
resulting in
\begin{equation}
m_\nu \simeq {-m_D^2 \over m_N},
\end{equation}
with mixing between $\nu_L$ and $N_R$ given by
\begin{equation}
\tan \theta_1 \simeq {m_D \over m_N} \simeq \sqrt{|m_\nu/m_N|}.
\end{equation}
Since $N_R$ does not have gauge interactions, the $3 \times 3$ mixing 
matrix linking the 3 neutrinos to the 3 charged leptons cannot be exactly 
unitary.  However, for $m_\nu \sim 1$ eV and $m_N \sim 1$ TeV, this violation 
of unitarity is of order $10^{-6}$, which is much too small to be observed. 
Note that lepton-number conservation is recovered in the limit $m_N \to 
\infty$.  [If neutrinos obtain Majorana masses directly through a Higgs 
triplet, thus doing without $N_R$, then there is no violation of unitarity 
to begin with.]

Consider now the idea of the $inverse$ seesaw mechanism: a situation is 
established where $m_\nu = 0$ because of a symmetry, which is then broken 
by a $small$ mass \cite{ww83,mv86,m87,gv89,gsv90,dv05,kk07}. In contrast to 
the canonical seesaw mechanism, lepton-number conservation is recovered 
here in the limit this small mass goes to zero.  The prototype model is to 
add a singlet Dirac fermion $N$, i.e. both $N_R$ and $N_L$, with lepton 
number $L=1$ per family to the SM.  The $3 \times 3$ mass matrix spanning 
($\bar{\nu}_L, N_R, \bar{N}_L$) is then given by
\begin{equation}
{\cal M}_{\nu,N} = \pmatrix{0 & m_D & 0 \cr m_D & \epsilon_R & m_N \cr 
0 & m_N & \epsilon_L},
\end{equation}
where $\epsilon_{L,R}$ are lepton-number violating Majorana mass terms. This 
is a natural extension of the famous seesaw $2 \times 2$ mass matrix of 
Eq.~(1), but it also has a clear symmetry interpretation, i.e. $\epsilon_{L,R}$ 
may be naturally small because their absence would correspond to the exact 
conservation of lepton number. [A linear combination of $\nu_L$ and $N_L$ 
would combine with $N_R$ to form a Dirac fermion, whereas its orthogonal 
combination would remain massless.]

Using $\epsilon_{L,R} << m_D, m_N$, the smallest mass eigenvalue of Eq.~(4) 
is then
\begin{equation}
m_\nu \simeq {m_D^2 \epsilon_L \over m_N^2},
\end{equation}
with mixing between $\nu_L$ and $N_L$ given by
\begin{equation}
\tan \theta_2 \simeq {m_D \over m_N} \simeq \sqrt{|m_\nu/\epsilon_L|}.
\end{equation}
Note first that the mixing between $\nu_L$ and $N_R$ remains negligible, 
i.e. $m_D \epsilon_L /m_N^2 \simeq m_\nu/m_D$.  More importantly, note that 
$m_N$ in Eq.~(3) is replaced by $\epsilon_L$ in Eq.~(6).  This means that 
$\theta_2$ is not constrained in the same way as $\theta_1$, and it can be 
bigger by orders of magnitude.  For example, let $m_\nu \sim 1$ eV and 
$\epsilon_L \sim 10$ keV, which is compatible with $m_D \sim 10$ GeV and 
$m_N \sim 1$ TeV, then the mixing is of order $10^{-2}$.  This dramatic 
change in possible unitarity violation means that it may be observable 
in future neutrino experiments \cite{bgw04,abfgl06,sz06,xz08,abf09,ahpz09,
moz09,abfl09,r09}.  If confirmed, it will be a big boost for the idea of 
the inverse seesaw.

If $\nu_e$ or $\nu_\mu$ mixes significantly with singlets, then the effective 
$G_F$ for $\mu \to e \nu_\mu \bar{\nu}_e$ would have to be redefined, and 
many precision electroweak measurements would be affected.  Thus only 
$\nu_\tau$ mixing is likely to be significant, affecting the unitarity of 
the third row of the neutrino mixing matrix linking $\nu_{e,\mu,\tau}$ to 
the mass eigenstates $\nu_{1,2,3}$.

In Eq.~(4), if $\epsilon_R = 0$ or relatively small, but $\epsilon_L$, renamed 
$m_L$, is very large, with $\epsilon_R, m_D << m_N^2/m_L$, then this is 
called the $double$ seesaw.  First, $N_R$ gets a medium large Majorana mass 
from $N_L$, i.e. $-m_N^2/m_L$.  For example, $m_N \sim 10^{9}$ GeV, $m_L \sim 
10^{15}$ GeV, then this seesaw mass is $\sim 1$ TeV.  Second, $\nu_L$ 
gets a small Majorana mass from $N_R$, i.e. $m_D^2 m_L/m_N^2$.  This idea 
is often used in models of grand unification.  In this case, the mixing 
between $\nu_L$ and $N_R$ is the same as in the canonical seesaw, and 
that between $\nu_L$ and $N_L$ is further suppressed.

A more recent proposal \cite{mrv05} is the $linear$ seesaw, i.e.
\begin{equation}
{\cal M}_{\nu,N} = \pmatrix{0 & m_D & m_2 \cr m_D & 0 & m_N \cr m_2 & m_N & 0},
\end{equation}
so that
\begin{equation}
m_\nu \simeq {-2 m_2 m_D \over m_N}.
\end{equation}
However, since both $N_R$ and $\bar{N}_L$ are singlets, they may be redefined 
by a rotation so that $m_2=0$.  Let $m_2/m_D = \tan \theta$, and $c=\cos 
\theta$, $s=\sin \theta$, $c_2=\cos 2 \theta$, $s_2 = \sin 2 \theta$, then
\begin{equation}
\pmatrix{1 & 0 & 0 \cr 0 & c & s \cr 0 & -s & c} \pmatrix{0 & m_D & m_2 \cr 
m_D & 0 & m_N \cr m_2 & m_N & 0} \pmatrix{1 & 0 & 0 \cr 0 & c & -s \cr 0 & s 
& c} = \pmatrix{0 & m_D/c & 0 \cr m_D/c & m_N s_2 & m_N c_2 \cr 0 & m_N c_2 & 
-m_N s_2},
\end{equation}
which is the same as Eq.~(4), resulting in
\begin{equation}
m_\nu \simeq (m_D^2/m_N^2)(-2 m_N m_2/m_D) = -2 m_2 m_D/m_N,
\end{equation}
which is identical to Eq.~(8) as expected.  On the other hand, if there is a 
symmetry beyond that of the SM which enforces Eq.~(7), then it may be 
considered on its own.

There is however another interesting variation (the $lopsided$ seesaw), 
which has not been discussed before. Let $\epsilon_{L,R}$ be renamed 
$m_{L,R}$, i.e. 
\begin{equation}
{\cal M}_{\nu,N} = \pmatrix{0 & m_D & 0 \cr m_D & m_R & m_N \cr 
0 & m_N & m_L},
\end{equation}
and do away with the notion of lepton number, then for
\begin{equation}
m_D << m_R, ~~~ {m_N^2 \over m_R} << m_L << m_N << m_R,
\end{equation}
the neutrino mass is again given by $m_\nu \simeq -m_D^2/m_R$ as in the 
canonical seesaw, and the mixing of $\nu_L$ and $N_R$ is again $m_D/m_R$, 
but now the mixing of $\nu_L$ and $N_L$ is $m_N m_D/m_L m_R < m_D/m_N$ which 
can be significant.  If $m_L$ is small enough, $N_L$ should be considered a 
sterile neutrino.  In that case, the violation of unitarity may show up also 
in the first two rows of the neutrino mixing matrix.  As an example, let 
$m_D \sim 1$ GeV, $m_R \sim 10^9$ GeV, $m_N \sim 10$ GeV, $m_L \sim 1$ keV, 
then $m_\nu \sim 1$ eV, and the mixing between $\nu_L$ and $N_L$ is $10^{-2}$.

If $m_L=0$ and $m_R$ is very large, it appears from Eq.~(7) at first sight 
that there are two seesaw masses, i.e. $-m_D^2/m_R$ and $-m_N^2/m_R$.  
However, since the determinant of ${\cal M}_{\nu,N}$ is zero in this case, 
the linear combination $(m_N \nu_L - m_D N_L)/\sqrt{m_N^2+m_D^2}$ remains 
massless.  If $m_L \neq 0$, i.e. the seesaw is lopsided, then there is no 
zero eigenvalue.

The neutrino mass matrices of Eqs.~(4) and (7) are also very suited for 
gauge extensions of the SM, such as $SU(3)_C \times SU(2)_L \times SU(2)_R 
\times U(1)_{B-L}$ and $SU(3)_C \times SU(2)_L \times U(1)_Y \times U(1)_X$, 
where $U(1)_X$ is orthogonal to $U(1)_Y$, an example of which is a linear 
combination of $U(1)_\chi$ and $U(1)_\psi$ in $E_6$ models.
As a concrete example, consider the lopsided seesaw and $U(1)_N$ of $E_6$ 
\cite{m96,km96,hmrs01,kmn06,ms07}.  Under the maximum subgroup $SU(3)_C \times 
SU(3)_L \times SU(3)_R$ of $E_6$, the charges of its fundamental 
\underline{27} representation under $U(1)_N$ are given by $Q_N = 6Y_L + T_{3R} - 
9Y_R$.  Hence $(u,d),u^c,e^c$ have $Q_N = 1$, $(\nu,e),d^c$ have $Q_N = 2$, 
and $N^c$ has $Q_N = 0$. To allow for quark and lepton masses, two Higgs 
scalar doublets
\begin{equation}
\Phi_1 = (\phi_1^0,\phi_1^-) \sim (1,2,-1/2,-3), ~~~ 
\Phi_2 = (\phi_1^+,\phi_1^0) \sim (1,2,1/2,-2)
\end{equation}
under $SU(3)_C \times SU(2)_L \times U(1)_Y \times U(1)_N$ are needed, with 
the Yukawa interactions
\begin{equation}
(u\phi_2^0 - d \phi_2^+) u^c, ~~ (d \phi_1^0 - u \phi_1^-) d^c, ~~ 
(e \phi_1^0 - \nu \phi_1^-) e^c, ~~ (\nu \phi_2^0 - e \phi_2^+) N^c.
\end{equation}
In the \underline{27} of $E_6$, there is another fermion singlet $S$ which 
has $Q_N =5$.  Consider now the breaking of $U(1)_N$ by the Higgs scalars 
$\chi_1 \sim -5$ and $\chi_2 \sim 10$.  The most general Higgs potential 
is given by
\begin{equation}
V_\chi = \sum_i \mu_i^2 \chi_i^\dagger \chi_i + {1 \over 2} \sum_{i,j} 
\lambda_{ij} (\chi_i^\dagger \chi_i)(\chi_j^\dagger \chi_j) + [\mu_{12} 
\chi_1^2 \chi_2 + H.c.],
\end{equation}
where $\lambda_{12} = \lambda_{21}$.  Let $\langle \chi_i \rangle = u_i$, then 
the conditions for $V_\chi$ to be at its minimum are
\begin{eqnarray}
&& u_1 [\mu_1^2 + \lambda_{11} u_1^2 + \lambda_{12} u_2^2 + 2\mu_{12} u_2] = 0,\\ 
&& u_2 [\mu_2^2 + \lambda_{12} u_1^2 + \lambda_{22} u_2^2] + \mu_{12} u_1^2 = 0.
\end{eqnarray}
A natural solution exists \cite{ms98,m01,glr09}, such that $u_2 << u_1$, i.e.
\begin{equation}
u_1^2 \simeq {-\mu_1^2 \over \lambda_{11}}, ~~~ u_2 \simeq {-\mu_{12} u_1^2 
\over \mu_2^2 + \lambda_{12} u_1^2}.
\end{equation}
If $\mu_{12} = 0$, there would be an extra global U(1) symmetry in $V_\chi$. 
Hence a small $\mu_{12}$ is natural, and $u_2 << u_1$ may be maintained. 
The $3 \times 3$ mass matrix spanning $(\nu, N^c, S)$ is then of the form 
desired, with $m_R$ an invariant mass, $m_N$ coming from $\chi_1$ and 
$m_L$ coming from $\chi_2$.

To summarize, it has been shown in this paper that a $3 \times 3$ realization 
of the seesaw mechanism has three distinguishable scenarios, resulting in the 
inverse (or linear) seesaw, the double seesaw, and the lopsided seesaw.  The 
last is a new proposal and may naturally be implemented in the $U(1)_N$ 
extension of the SM.

This work was supported in part by the U.~S.~Department of Energy under 
Grant No.~DE-FG03-94ER40837.

\baselineskip 18pt
\bibliographystyle{unsrt}

\end{document}